\documentclass[%
superscriptaddress,
showpacs,preprintnumbers,
amsmath,amssymb,
aps,
prl,
twocolumn
]{revtex4-1}
\usepackage{ulem}
 \usepackage{graphicx}
\usepackage{dcolumn}
\usepackage{bm}
\usepackage{color}

\begin {document}

\title{
Diffusion of intrinsically disordered proteins within viscoelastic membraneless droplets
}

\author{Fuga~Watanabe}
\affiliation{Department of System Design Engineering, Keio University, Yokohama, Kanagawa 223-8522, Japan}

\author{Takuma~Akimoto}
\affiliation{Department of Physics, Tokyo University of Science, Noda, Chiba 278-8510, Japan}

\author{Robert~B.~Best}
\affiliation{Laboratory of Chemical Physics, National Institute of Diabetes and Digestive and Kidney Diseases, National Institutes of Health, Bethesda, USA}

\author{Kresten~Lindorff-Larsen}
\affiliation{Linderstr{\o}m-Lang Centre for Protein Science, Department of Biology, University of Copenhagen, DK-2200 Copenhagen N, Denmark}

\author{Ralf~Metzler}
\affiliation{Institute of Physics \& Astronomy, University of Potsdam, 14476 Potsdam-Golm, Germany}
\affiliation{Asia Pacific Centre for Theoretical Physics, Pohang, 37673, Republic of Korea}

\author{Eiji~Yamamoto}
\email{eiji.yamamoto@sd.keio.ac.jp}
\affiliation{Department of System Design Engineering, Keio University, Yokohama, Kanagawa 223-8522, Japan}




\begin{abstract}
In living cells, intrinsically disordered proteins (IDPs), such as FUS and DDX4, undergo phase separation, forming biomolecular condensates.
Using molecular dynamics simulations, we investigate their behavior in their respective homogenous droplets.
We find that the proteins exhibit transient subdiffusion due to the viscoelastic nature and confinement effects in the droplets. 
The conformation and the instantaneous diffusivity of the proteins significantly vary between the interior and the interface of the droplet, resulting in non-Gaussianity in the displacement distributions.
This study highlights key aspects of IDP behavior in biomolecular condensates.
\end{abstract}

\maketitle

Multivalent proteins and nucleic acids form biomolecular condensates, which are so-called membraneless organelles (MLOs), through phase separation~\cite{HymanWeberJuelicher2014, BananiLeeHymanRosen2017, PappuCohenDarFaragKar2023}.
A primary driving force for phase separation is the sequence-specific multivalent interactions among the components.
Notably, intrinsically disordered proteins (IDPs), due to their flexibility and low-complexity sequences, play a crucial role in the formation of these condensates~\cite{ProtterRaoVanTreeckLinMizoueRosenParker2018, WangChoiHolehouseLeeZhangJahnelMaharanaLemaitrePozniakovskyDrechselPoserPappuAlbertiHyman2018}. 
The organization of MLOs has been identified as contributing to essential biological processes, such as gene regulation, biochemical reactions, signal transduction, and molecular protection as a stress response.
These functions depend on the unique physical properties of the condensates.
The presence of flexible IDPs render the interior of the condensate highly dynamic and complex, exhibiting liquid-like properties with viscoelasticity~\cite{BrangwynneEckmannCoursonRybarskaHoegeGharakhaniJuelicherHyman2009, Elbaum-GarfinkleKimSzczepaniakChenEckmannMyongBrangwynne2015, JawerthIjaviRuerSahaJahnelHymanJuelicherFischerFriedrich2018, JawerthFischerFriedrichSahaWangFranzmannZhangSachwehRuerIjaviSahaMahamidHymanJuelicher2020, AlshareedahMoosaPhamPotoyanBanerjee2021}.

Several studies have investigated the dynamic property of proteins within the condensates using experimental and computational methods such as pulsed field gradient NMR~\cite{MurthyDignonKanZerzeParekhMittalFawzi2019}, fluorescence recovery after photobleaching (FRAP)~\cite{BurkeJankeRhineFawzi2015}, single-molecule fluorescence microscopy~\cite{Elbaum-GarfinkleKimSzczepaniakChenEckmannMyongBrangwynne2015, FisherElbaum-Garfinkle2020, KamagataIwakiHazraKanbayashiBanerjeeChibaSakomotoGaudonCastaingTakahashiKimuraOikawaTakahashiLevy2021, KamagataIwakiKanbayashiBanerjeeChibaGaudonCastaingSakomoto2022} and molecular dynamics (MD) simulations~\cite{ZhengDignonJovicXuRegyFawziKimBestMittal2020, BenayadBuelowStelzlHummer2021, TejedorGaraizarRamirezEspinosa2021, TejedorSanchez-BurgosEstevez-EspinosaGaraizarCollepardo-GuevaraRamirezEspinosa2022, RanganathanShakhnovich2022}.
The diffusivity of proteins within condensates is significantly reduced compared to that in bulk and can be tuned by sequence variation and length.
It has also been highlighted that the interface of biomolecular condensates acts as a key element in driving pathological protein aggregation~\cite{ShenChenWangRuggeriAimeWangQamarEspinosaGaraizarCollepardoGuevaraWeitzVigoloKnowles2023} and molecular exchange between the condensates and bulk~\cite{BoHubatschBauermannWeberJuelicher2021}.
A detailed exploration of the intricate diffusion process of IDPs using physical models could shed light on the properties pivotal for cellular processes.
However, a comprehensive understanding of the spatiotemporal conformation and dynamics of IDPs within the condensates remains elusive at the molecular scale.

Here, we perform large-scale MD simulations of spherical droplets formed by mono IDPs to investigate the diffusive dynamics of proteins within the droplets.
We demonstrate that proteins exhibit the anomalous diffusion with fluctuating diffusivity attributed to the viscoelastic environment of the droplet.
Furthermore, we show that diffusivity varies between the interior and the interface of the droplets, resulting in a non-Gaussian distribution of displacement.

\begin{figure}[tb]
\centering
\includegraphics[width = 70 mm,bb= 0 0 231 270]{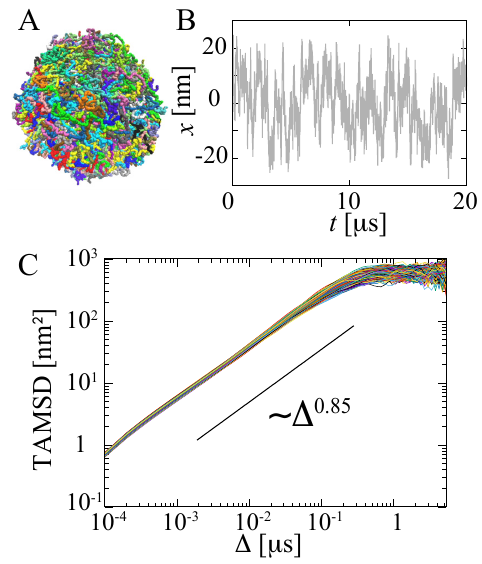}
\caption{Diffusion of FUS-LCD within a droplet.
(A)~The snapshot of the droplet, with a radius of approximately $ 20 \, \mathrm{nm}$, formed by 1,000 molecules of FUS-LCD.
Each color in the snapshot represents a different protein.
(B)~Relative $x$-coordinate of a protein with respect to the droplet's COM.
(C)~The TAMSDs for the FUS-LCD proteins within the droplet, where different colors represent different proteins.}
\label{fig1}
\end{figure}

{\it Subdiffusive motion of proteins within the droplet}--We performed coarse-grained (CG) MD simulations of droplets formed by 1,000 IDPs, well-studied Fused in Sarcoma low-complexity domains (FUS-LCD), using a residue-revel CG model~\cite{DignonZhengKimBestMittal2018, RegyThompsonKimMittal2021, TeseiSchulzeCrehuetLindorff-Larsen2021} (see Fig.~\ref{fig1}A and details in SI).
Within the simulation timescales of $20 \, \mathrm{\mu s}$, proteins diffuse between the interior and the interface of the droplet several times (Fig.~Fig.~\ref{fig1}B).
To characterize the diffusive dynamics of proteins, the time-averaged mean squared displacement (TAMSD) 
\begin{equation}
    \overline{\delta^{2}_{i}(\Delta;t)} = \frac{1}{t-\Delta}\int_{0}^{t-\Delta} \{\vec{r}_i(t' + \Delta ) - \vec{r}_i(t')\}^2 dt'
\end{equation}
of proteins was calculated, where $t$ is the observation time, $\Delta$ is the lag time, and $\vec{r}_i$ is the relative coordinates of the center of mass (COM) of the $i$th protein obtained by subtracting the COM position of the droplet.
In what follows, we exclusively analyzed proteins that reside within the droplet during the observation time.
Figure~\ref{fig1}C shows that TAMSDs of each protein exhibit sub-linear increase, $\overline{\delta^{2}_{i}(\Delta;t)} \propto \Delta^{\alpha}$, with $\alpha \sim 0.85$ within the droplets.
After $1 \, \mathrm{\mu s}$ the TAMSDs show a plateau, indicative of the spatial confinement of proteins within the droplet, where the plateau value of the TAMSDs corresponds to the square of the droplet radius, approximately $20 \, \mathrm{nm}$.
Note that the ensemble-averaged MSD is consistent with the TAMSDs, indicating that the diffusion process is ergodic (see Fig.~S1).
The subdiffusive behavior is distinct from the Brownian motion of an isolated protein in a solution, where the TAMSD increases linearly without saturation.
The subdiffusive motions are often observed in the molecular diffusion within viscoelastic media, e.g. polymer solutions~\cite{MasonWeitz1995, MasonGanesanZantenWirtzKuo1997} and lipid membranes~\cite{AkimotoYamamotoYasuokaHiranoYasui2011, JeonMonneJavanainenMetzler2012, YamamotoAkimotoYasuiYasuoka2014, JeonJavanainenMartinez-SearaMetzlerVattulainen2016}.
We also confirmed similar subdiffusion for another protein (DDX4), under different temperature conditions, and with different model parameters (see Fig.~S2).

\begin{figure}[tb]
\centering
\includegraphics[width = 70 mm,bb= 0 0 170 256]{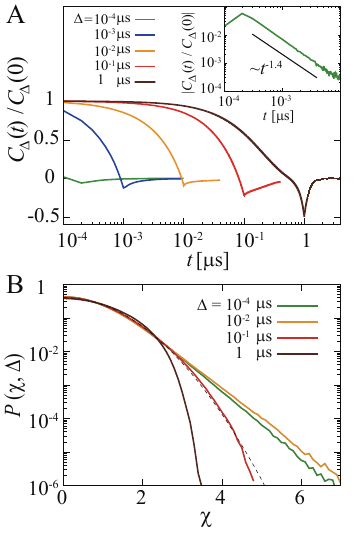}
\caption{Anticorrelated motion at short times and non-Gaussianity of protein movements within droplets.
(A)~Normalized DAF of the protein with different lag time $\Delta$.
The inset shows the log-log plot for $\Delta = 10^{-4} \, \mathrm{\mu s}$.
The black line indicates the theory of FBM.
(B)~The propagators as a function of the normalized position, defined by $x / \sigma \equiv \chi$, for different lag time $\Delta$, where $\sigma$ is the standard deviation corresponding to each $\Delta$.
The black dashed line represents a Gaussian distribution with unit variance.}
\label{fig2}
\end{figure}

{\it Anticorrelated motion and non-Gaussian distribution of protein movements}--Fractional Brownian motion (FBM)~\cite{Kolmogorov1940, MandelbrotVan1968} is a fundamental model of anomalous diffusion characterized by Brownian motion driven by long-term correlated noise.
If the noise exhibits anticorrelation, the TAMSD shows a sublinear increase~\cite{MetzlerJeonCherstvyBarkai2014}.
Figure~\ref{fig2}A shows the normalized form of the displacement autocorrelation function (DAF) of proteins,
\begin{equation}
    C_{\Delta}(t) = \frac{\langle (\vec{r}(t+\Delta) - \vec{r}(t))(\vec{r}(\Delta)-\vec{r}(0)) \rangle}{{\Delta}^2},
\end{equation}
where $\langle \cdot \rangle$ indicates ensemble average and $\Delta = 0.1 \, \mathrm{ns}$.
The transition of the DAF from positive to negative values indicates anticorrelated motion.
The dip in the DAFs signifies the degree of anticorrelation.
Viscoelastic effects are significant on a short timescale and diminish gradually over $\Delta = 10^{-2} \, \mathrm{\mu s}$.
Conversely, boundary confinement effects become noticeable around $\Delta = 10^{-1} \, \mathrm{\mu s}$ and are more pronounced at $\Delta = 1 \, \mathrm{\mu s}$, a dip of sharp peak of $-0.5$~\cite{WeberThompsonMoernerSpakowitzTheriot2012}, where the TAMSDs reach a plateau.
Moreover, the normalized DAF exhibits a power-law decay, although it slightly deviates from the free FBM~\cite{BurovJeonMetzlerBarkai2011}, $C_{\Delta}(t) / C_{\Delta}(0) \propto t^{\alpha - 2}$, where $\alpha$ is the same exponent as in $\overline{\delta^{2}_{i}(\Delta;t)} \propto \Delta^{\alpha}$ (Fig.~\ref{fig2}A).
As expected from the FBM with a DAF exponent of $-1.4$, the MSD should increase sublinearly with $\alpha = 0.6$.
The slight increase in the MSD exponent might be attributed to factors such as protein entanglements, conformational changes, etc.

To examine the Gaussianity of the displacement, we calculated a normalized propagator $P(\chi, \Delta)$, which represents the probability density function (PDF) of the displacement that the particle diffuses within a lag time $\Delta$.
Since no systematic differences were observed in $x, y, z$ directions, we combined all increments from these directions into a single set denoted as $\chi$, i.e. displacements calculated as $x(t+\Delta) - x(t)$, $y(t+\Delta) - y(t)$, and $z(t+\Delta) - z(t)$ are collectively combined and represented as $\chi$.
In the case of a normal FBM, the normalized propagator is expected to be Gaussian and consistent for different $\Delta$.
Figure \ref{fig2}B shows that the propagators of proteins in the droplet exhibit non-Gaussianity with different $\Delta$ values.
On a short time scale, until $\Delta = 0.1 \, \mathrm{\mu s}$, there are significant deviations from the Gaussian distribution in the tails of the distribution.
Such non-Gaussianity has been observed in the diffusion process with fluctuating diffusivity~\cite{WangKuoBaeGranick2012, ChubynskySlater2014, HeSongSuGengAckersonPengTong2016, CherstvyMetzler2016, ChechkinSenoMetzlerSokolov2017}, i.e., individual proteins within the droplet diffuse non-uniformly.
Conversely, on a long time scale, the propagator is close to the Gaussian distribution.
The $\Delta = 0.1 \, \mathrm{\mu s}$ is the same timescale at which the TAMSD converges to a plateau (see Fig.~\ref{fig1}C), which again indicates the confinement effect in the droplet.

\begin{figure}[tb]
\centering
\includegraphics[width = 70 mm,bb= 0 0 218 155]{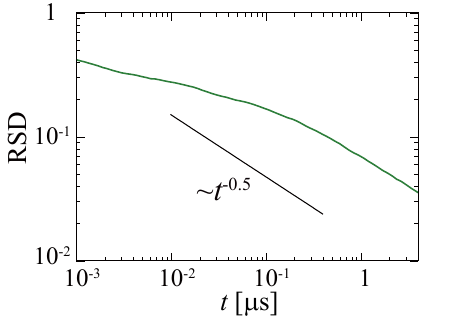}
\caption{RSD of TAMSDs.
The black solid line is a $t^{-0.5}$ scaling, shown for reference.}
\label{fig3}
\end{figure}

{\it Fluctuating diffusivity of proteins within the droplet}--To quantitatively evaluate the fluctuation of the diffusivity, we calculated the relative standard deviation (RSD) of TAMSD~\cite{UneyamaAkimotoMiyaguchi2012, YamamotoAkimotoKalliYasuokaSansom2017}, the square root of the ergodicity breaking parameter ~\cite{BurovJeonMetzlerBarkai2011},
\begin{equation}
    \rm{RSD} = \frac{\sqrt{\langle \overline{\delta^{2}(\Delta;\it{t})}^2\rangle - \langle \overline{\delta^{2}(\Delta;\it{t})}\rangle^2}}
    {\langle \overline{\delta^{2}(\Delta;\it{t})} \rangle}.
\end{equation}
In diffusion following a Gaussian process such as Brownian motion, the RSD decays as $t^{-0.5}$, but for non-ergodic processes such as CTRW~\cite{HeBurovMetzlerBarkai2008, MetzlerJeonCherstvyBarkai2014} or with significant fluctuation in the diffusivity~\cite{AkimotoYamamoto2016, MiyaguchiAkimotoYamamoto2016, AkimotoYamamoto2016a}, the RSD does not decay to zero even if observed for a long time.
In polymeric solutions, the RSD changes from non-Gaussian fluctuations (RSD decays at $t^{-\alpha}(0<\alpha<0.5)$) to Gaussian fluctuations (RSD decays at $t^{-0.5}$) because of polymer entanglement, and the crossover time corresponds to the longest relaxation time~\cite{UneyamaAkimotoMiyaguchi2012, UneyamaMiyaguchiAkimoto2015}.
The RSD of FUS-LCD exhibits slow decay at short timescales, then a crossover to Gaussian fluctuations (see Fig.~\ref{fig3}).

\begin{figure*}[tb]
\centering
\includegraphics[width = 150 mm,bb= 0 0 361 261]{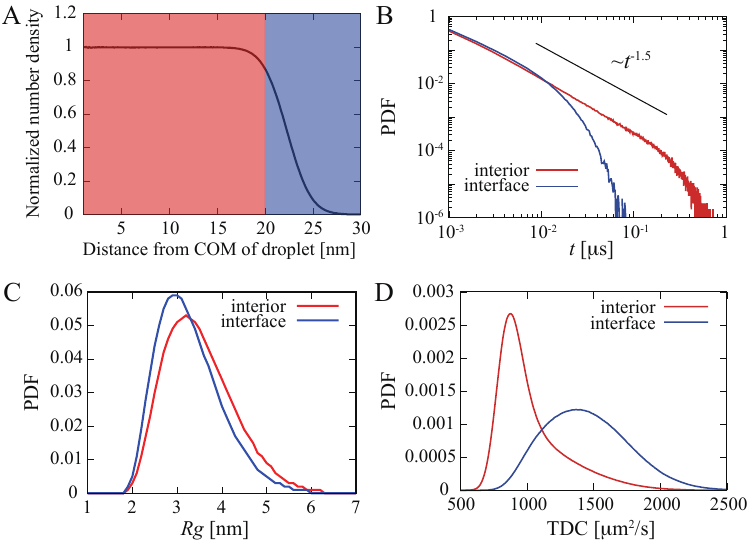}
\caption{The difference in conformation and diffusivity of proteins at the interior and interface of the droplet.
(A)~The normalized radial number density of proteins within the droplet.
The red and blue regions correspond to the interior and the interface of the droplet, respectively.
(B)~The PDF of residence times for proteins at the interior and the interface of the droplet.
(C)~The PDF of $R_g$ and (D)~TDC.}
\label{fig4}
\end{figure*}

{\it Conformation and diffusivity of proteins at the interior and the interface of the droplets}--From the radial density profile, we define the boundary between the interior and the interface of the droplet by $20 \,\mathrm{nm}$ (see Fig.~\ref{fig4}A).
The residence time distributions for the interior and the interface of the droplet exhibit a power-law distribution with an exponential cutoff $P(t) \propto t^{-\beta} \exp{(-t/\tau)}$ (see Fig.~\ref{fig4}B).
The power-law exponent $\beta = -1.5$ is consistent with the first-passage time distribution of one-dimensional Brownian motion~\cite{KaratzasKaratzasShreveShreve1991, GodecMetzler2016}.
The cutoff of the residence time of protein within the droplet interior ($\sim 0.2  \, \mathrm{\mu s}$) is comparable to the crossover time of the RSD, suggesting that the non-Gaussian fluctuations are due to transitions between residence in the interior and the interface.

The regions either in the interior or in the interface of the droplet can influence the structure and dynamics of proteins.
The distribution of the radius of gyration $R_g$ has a singular peak, and proteins in the interior of the droplet exhibit a slightly larger conformation than those at the interface (see Fig.~\ref{fig4}C).
This is attributed to the hydrophobicity inside the droplet.
To compare diffusive dynamics, we calculated the temporary diffusion coefficient (TDC)~\cite{YamamotoAkimotoKalliYasuokaSansom2017, AkimotoYamamoto2017, YamamotoAkimotoMitsutakeMetzler2021},
\begin{equation}
D(t^*) = \frac{1}{2d\Delta(t-\Delta)}\int_{t^*}^{t^* + t - \Delta} \Big \{ \vec{r}(t'+ \Delta) - \vec{r}(t')\Big \}^2 dt', 
\end{equation}
where $d$ is the spatial dimension, $\Delta$ is $0.1  \, \mathrm{ns}$, and $t$ is $10  \, \mathrm{ns}$.
In the TDC analysis, we consider $D(t^*)$ for a protein interior or interface regions at $t = t^*$.
The TDC distribution implies that proteins in the interface region diffuse faster than those in the interior of the droplet.
This observation is due to the reduced density and the smaller size of proteins at the interface compared to those within the droplet (Fig.~\ref{fig4}D). 
The movement between the interior of the droplet and its surface leads to significant fluctuations in diffusivity (Figs.~\ref{fig2}B and~\ref{fig3}).

{\it Universality of fluctuating diffusivity in droplet}--The universal nature of the diffusive protein behavior within spherical droplets is underscored by MD simulations performed at different temperatures ($320  \, \mathrm{K}$), with different protein N-terminal disordered regions of DDX4 (DDX4-DR), and using different force fields (see Fig.~S2--S7).
Qualitatively, no clear differences are observed between FUS-LCD and DDX4-DR.
When droplets form from 1,000 proteins, the density inside the DDX4-DR droplet is lower than that of the FUS-LCD droplet (Fig.~S6).
Additionally, because the residue length of DDX4-DR is greater than that of FUS-LCD, the droplet size of DDX4-DR becomes larger than that of FUS-LCD.
This results in slightly faster diffusivity and a longer relaxation time for the fluctuating diffusivity of DDX4-DR compared to those of FUS-LCD (see Figs.~3, S5, S7).
If the observation time scale is sufficiently longer than the scale at which proteins significantly diffuse between the droplet's surface and its interior, diffusive behavior might be more determined by the size of the protein and the density of the droplet than by the sequence-specific interactions of IDPs.

{\it Discussion}--
In summary, we performed MD simulations to investigate individual protein diffusion within protein droplets.
We revealed subdiffusion within the droplets due to their viscoelastic nature.
The diffusivity of proteins differs significantly between the interior and the interface of the droplet.
The relaxation time of these fluctuations is equivalent to the residence time of the proteins within the interior of the droplets.
Information on protein diffusion within the droplet and residence time in interior and interface regions will lead to an understanding of cell functions such as chemical reactions, transport of nanoparticles, and biomolecular storage using the droplet as a medium.

Performing large-scale MD simulations at the atomic level that are comparable to experimental spatiotemporal scales remains challenging~\cite{ZhengDignonJovicXuRegyFawziKimBestMittal2020}.
Residue-level CG-MD simulations with an implicit solvent are a unique and powerful tool to explore molecular phenomena over tens to hundreds of nanometers and tens of microseconds~\cite{DignonZhengKimBestMittal2018, JosephReinhardtAguirreChewRussellEspinosaGaraizarCollepardo-Guevara2021}.
However, the diffusion coefficient of proteins within the droplets, as simulated using the CG models, was three orders of magnitude larger than the experimental values~\cite{BurkeJankeRhineFawzi2015, MurthyDignonKanZerzeParekhMittalFawzi2019,KamagataIwakiHazraKanbayashiBanerjeeChibaSakomotoGaudonCastaingTakahashiKimuraOikawaTakahashiLevy2021}.
This implies that simulations of $20 \, \mathrm{\mu s}$ correspond to real-time durations of $20 \, \mathrm{ms}$ (see Fig.~S8).
Fast dynamics of the residue-revel CG models have been also reported in other studies~\cite{TejedorGaraizarRamirezEspinosa2021, TejedorSanchez-BurgosEstevez-EspinosaGaraizarCollepardo-GuevaraRamirezEspinosa2022}, as these models were parameterized to reproduce the structural and phase behavior rather than the dynamic properties~\cite{RegyThompsonKimMittal2021, JosephReinhardtAguirreChewRussellEspinosaGaraizarCollepardo-Guevara2021}.
The higher diffusivity of residue-level CG models with implicit solvent is due to the absence of factors such as side-chain entanglements and hydrodynamic interactions.
One method to match the simulated diffusion coefficients with experimental values is by adjusting the friction coefficient in the Langevin equation.
Increasing the friction coefficient by factors of 10 to 1,000 resulted in a decrease of the diffusion coefficient from 200 to $4.0 \, \mathrm{\mu m^2 /s}$, respectively, eventually reaching the same order of magnitude as the experimental values.
Note that increasing the friction coefficient does not alter the structural property of $R_g$ and the density of the droplets, and the physical mechanism underlying the diffusion process remains qualitatively consistent.

\subsection*{Acknowledgments}
This work was supported by JSPS KAKENHI Grant Number 21H05728 and JST PRESTO Grant Number JPMJPR22EE, Japan.
We thank computer FUGAKU(ID: hp230327) for providing computing resources for this work.
R.M. thanks the German Science Foundation (DFG, grant no. ME1535/12-1) for support.


%

\end{document}